\documentclass[conference,peer-reviewed]{aesconf} 
\usepackage{hyperref}
\usepackage{placeins}
\usepackage{color, colortbl}
\usepackage{booktabs}
\usepackage{url}
\usepackage [autostyle, english = american]{csquotes}

\newcommand{\cmmnt}[1]{}

\usepackage[utf8]{inputenc}

\usepackage{microtype}

\usepackage[numbers,square]{natbib}

\definecolor{Red}{rgb}{0.73,0.56,0.56}
\definecolor{Blue}{rgb}{0.52,0.80,0.98}
\definecolor{Green}{rgb}{0.59,0.98,0.59}
\definecolor{Yellow}{rgb}{1,1,0.87}
\definecolor{Orange}{rgb}{0.91,0.58,0.47}

\definecolor{Red2}{rgb}{0.803921568627451, 0.3607843137254902, 0.3607843137254902}
\definecolor{Blue2}{rgb}{0.27450980392156865, 0.5098039215686274, 0.7058823529411765}
\definecolor{Green2}{rgb}{0.19607843137254902, 0.803921568627451, 0.19607843137254902}
\definecolor{Yellow2}{rgb}{1.0, 1.0, 0.0}
\definecolor{Orange2}{rgb}{1.0, 0.6470588235294118, 0.0}

\title{The Intrinsic Memorability of Everyday Sounds}

\author[]{David B. Ramsay$^{*}$}
\author[]{Ishwarya Ananthabhotla\thanks{equal contribution. The authors would like to thank the AI Grant for their financial support of this work.}}
\author[]{Joseph A. Paradiso}

\affil[]{Responsive Environments, MIT Media Lab}

\correspondence{Ishwarya Ananthabhotla}{Ishwarya@mit.edu}

\lastnames{Ramsay, Ananthabhotla, and Paradiso}

\shorttitle{Intrinsic Memorability of Sound}

\savebox{\AEStop}{%
	\begin{minipage}{\textwidth}%
		\rule{\textwidth}{1.5pt}\\%
		\\%
		\begin{minipage}[c][\iftoggle{convention}{3.2cm}{3.7cm}][t]{0\textwidth}%
			\includegraphics[width=20mm]{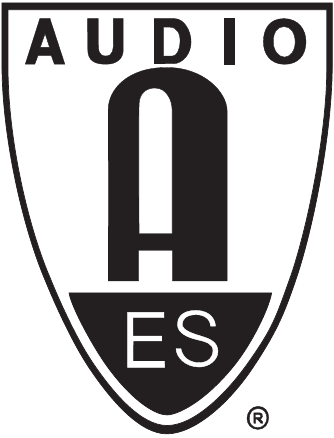}%
		\end{minipage}%
		\begin{minipage}{\textwidth}%
			\sffamily%
			\begin{center}%
				\LARGE Audio Engineering Society\\%
				\iftoggle{e_brief}{%
				\hspace{3mm}\fontsize{36}{38pt}\selectfont Convention e-Brief \AESEBriefNumber\\%
				}{%
				\iftoggle{convention}{%
				\fontsize{36}{38pt}\selectfont Convention Paper\\%
				}{%
				\fontsize{36}{38pt}\selectfont Conference Paper\\%
				}}%
				\vspace{0.2cm}%
				\large Presented at the \AESConferenceNumber \iftoggle{convention}{Convention\\}{Conference on\\}%
				\iftoggle{convention}{}{\AESConferenceTopic\\}%
				\AESConferenceDate, \AESConferenceLocation%
			\end{center}%
		\end{minipage}\\%
		\vspace{0.2cm}\\%
		\begin{minipage}{\textwidth}%
			\rmfamily\itshape\small	\AESLegalText%
		\end{minipage}\\%
		\\%
		\rule{\textwidth}{1.5pt}%
	\end{minipage}%
}

\begin{document}

\twocolumn[
\maketitle
\begin{onecolabstract}
Our aural experience plays an integral role in the perception and memory of the events in our lives.  Some of the sounds we encounter throughout the day stay lodged in our minds more easily than others; these, in turn, may serve as powerful triggers of our memories.  In this paper, we measure the memorability of everyday sounds across 20,000 crowd-sourced aural memory games, and assess the degree to which a sound's memorability is constant across subjects.  We then use this data to analyze the relationship between memorability and acoustic features like harmonicity, spectral skew, and models of cognitive salience; we also assess the relationship between memorability and high-level features with a dependence on the sound source itself, such as its familiarity, valence, arousal, source type, causal certainty, and verbalizability.  We find that (1) our crowd-sourced measures of memorability and confusability are reliable and robust across participants; (2) that the authors' measure of collective causal uncertainty detailed in our previous work, coupled with measures of visualizability and valence, are the strongest individual predictors of memorability; (3) that acoustic and salience features play a heightened role in determining "confusability" (the false positive selection rate associated with a sound) relative to memorability, and that (4), within the framework of our assessment, memorability is an intrinsic property of the sounds from the dataset, shown to be independent of surrounding context.  We suggest that modeling these cognitive processes opens the door for human-inspired compression of sound environments, automatic curation of large-scale environmental recording datasets, and real-time modification of aural events to alter their likelihood of memorability.

\end{onecolabstract}
]

\saythanks
\section{Introduction}
For a sound to enter our memory, it is first unconsciously processed by a change-sensitive, gestalt neural mechanism before passing through a conscious filtering process \cite{snyder2017recent,winkler2009modeling,inui2010non}.  We then encode this auditory information via a complex and variable process; frequently we abstract our experiences into words, though we also utilize phonological-articulatory, visual/visuospatial, semantic, and echoic memory \cite{buchsbaum2005human,vaidya2002evidence}.  
Different types of memory may also drive more visceral forms of recollection and experience; non-semantic memory, for example, may underpin powerful recollection and nostalgia experiences similar to those reported with music \cite{jancke2008music}.

In this work, we map out the features of everyday sounds that drive their memorability using an auditory memory game.  As a recall experiment, we hypothesize that it can provide useful insights into cognitive models for auditory capture and curation.  Additionally, we design the task such that it is beyond the capacity of our working and echoic memory and engages long-term memory cognitive processes \cite{ma2014changing,loaiza2015long}.  
With this work we hope to illuminate the role of top-down features -- imageability, emotionality, causal certainty, and familiarity -- in auditory memory. Using state-of-the-art cognitive saliency models, we also explore the relative importance of low-level acoustic descriptors against high-level conceptual ones for memory formation.  To our knowledge, this is the first general treatment of auditory memorability that combines low-level auditory salience models with multi-domain, top-down cognitive gestalt features.  This work enables more accurate models of auditory memory, and represents a step toward cognitively-inspired compression of everyday sound environments, automatic curation of large-scale environmental recording datasets, and real-time modification of aural events to alter their likelihood of memorability.

\section{What Influences Memorability?}
Many factors influence the cognitive processes underlying human aural processing and storage. Research shows a complicated interdependence between attention, acoustic feature salience, source concept salience, emotion, and memory; furthermore, verbal, pictorial, and phonological-articulatory mnemonics can have a significant impact on sound recall tasks.  

Neuroscience research supports the idea that gestalt auditory pre-processing is followed by attentive filtering prior to conscious perception \cite{snyder2017recent,winkler2009modeling}.  These gestalt representations incorporate both 'bottom-up' and 'top-down' processes -- sounds that are contextually novel only based on their acoustic features, as well as sounds that are only conceptually novel, lead to distinct and measurable variations in unconscious event-related potentials (ERPs) \cite{schirmer2011perceptual}.  These data motivate the need to incorporate high-level conceptual features and low-level acoustic features relative to a sound context for even simple models of auditory processing, attention, and memory.  

The stored gestalt representation of the current sound context -- necessary to explain change-driven ERPs -- can be thought of as the first stages of auditory memory \cite{inui2010non}.  This immediate store, known as 'echoic memory', starts decaying exponentially by 100ms after a sound onset \cite{lu1992behavioral}.  Measurements of ERPs suggest immediate storage of rhythmic stimuli on the order of 100 ms with a resolution as low as 5 ms \cite{nishihara2014echoic}; other studies have shown this immediate store is complimented with an additional echoic mechanism that lasts several seconds \cite{cowan1984short,alain1998distributed}.  On these time-scales, our auditory system compresses its perceptual representation of textures based on time-averaged statistics \cite{mcdermott2013summary}.

These principles have been used to design 'bottom-up' cognitive saliency models \cite{kayser2005mechanisms,delmotte2012computational}. While other time-averaged low-level features have been used to quantify sound similarity \cite{richard2013overview}, saliency models are now common in practical applications \cite{schauerte2011multimodal,kalinli2009saliency}.  Although the above work does not include higher-level gestalt processing, a few researchers have successfully combined low-level saliency modeling with a focused, task-specific top-down cognitive model \cite{kalinli2008combining,kalinli2009prominence,marchegiani2012top}.  These models aren't designed to generalize outside of their domain, however.

In general, high-level 'top-down' features have been an area of intense study that begins in the 1950's, when Colin Cherry demonstrated that his subjects noticed their name -- and no other verbal content -- spoken by a secondary speaker in a shadowing test \cite{cherry1953some}.  Besides the ongoing work in verbal auditory processing, research into non-verbal stimuli and auditory memory also provides us with insight into the role of conceptual abstractions in modulating attention and memory in a more general sense.  

One such abstraction -- emotionality -- is known to have a powerful effect on cognitive processing and memory formation \cite{ledoux1994emotion}. For music, recall has been shown to improve with positive valence, high arousal sound events \cite{jancke2008music}, though recent research has called the significance of arousal into question \cite{eschrich2008unforgettable}.  In noise pollution research, the high-level perception of human activity is considered 'pleasant' (more positively valenced) regardless of low-level acoustic features \cite{dubois2006cognitive}.  In general, the emotional impact of a sound is correlated with the clarity of its perceived source \cite{OURSELVES}, though sounds can have emotional impact even without a direct mapping to an explicit abstract idea \cite{quirin2009nonsense}.   

For recognition and recall memory exercises, verbalizing a sound or naming a sound (both of which may engage phonological-articulatory motor memory) is the most common and successful strategy \cite{bartlett1977remembering}. This semantic abstraction has overshadowing effects, though; verbal descriptions can distort recollection of the sound itself, degrading recognition performance without altering confidence \cite{mitchell2011remembering}.  Some researchers specifically isolate and study echoic memory, separating it from naming (a process that doesn't involve outward verbalization) using homophonic sound sources to ensure subjects are not relying internally on a naming mechanism \cite{conrad1972developmental}.  In everyday life, though, we naturally rely on a complex mixture of echoic (perceptual), phonological-articulatory (motor), verbal (semantic), and visual memory \cite{buchsbaum2005human,vaidya2002evidence}.

In our previous work \cite{OURSELVES} we curated a large dataset of everyday sounds to include high-level features that may influence a sound's memorability; most notably, its causal certainty (the degree to which a sound implies a clear, unambiguous source, denoted as $H_{cu}$), the implied source itself as determined by crowd-sourced workers, and its acoustic features.  We also collected ratings for the valence and arousal of each sound, its familiarity, and how easily it conjures a mental image (features that have strong correlations with $H_{cu}$).  Combined, these data give us insight into a sample's emotionality, as well as the ease by which it can be stored in semantic or visual memory.  Embeddings based on location relationships of sound sources ('at-location', 'located-near') give us additional insight into conceptual distinctness of a sound compared to the contextualizing sounds of a soundscape, and can serve as a first-order proxy for ecological exposure.

In this paper, we explore the relationship between both low-level acoustic features and high-level conceptual features with the memorability of sound, in a context that engages long-term memory processes.  We hope that a thorough analysis of the major low- and high-level features from the literature might lay the foundation for a practical, generalized model of auditory memory.

We set out to test a few important hypotheses, namely:

\textbf{1.} The cognitive processing of sound is similar enough across people that trends in recall across sound samples will be measurable and robust across users.

\textbf{2.} Higher-level gestalt features will be most predictive of successful recall performance.  We see from the literature that naming and emotionality have very strong effects in similar tasks-- we expect sounds with low $H_{cu}$ (easy to name their source) and strong valence/high arousal to be the most memorable.  High $H_{cu}$ (uncertain) sounds elicit weaker emotions, reinforcing this effect.  

\textbf{3.} Low-level acoustic feature information will marginally predict memory performance.  Gestalt features are not easily mapped to low level feature space (and we expect gestalt features to dominate); however, the literature suggests a measurable, second-order contribution from low-level perceptual saliency modeling.

\textbf{4.} The likelihood of a sound eliciting a false memory will be best predicted by its conceptual familiarity as well as by low-level acoustic features.  

\textbf{5.} The context a sound is presented against will have a marginal but measurable impact on whether it is recalled.  In other words, we expect emotional and unambiguous sounds to be the most memorable regardless of presentation, but when a sound stands out against the immediately preceding sounds, we hypothesize that it will be slightly more memorable.

\begin{figure*}[h]
\centering
\resizebox{0.88\textwidth}{!}{
\begin{tabular}{|c|c|}
  \hline
  \textbf{Vomit} & \textbf{Marketplace} \\
  \hline
  
  \begin{minipage}{.47\textwidth}
  	\vspace{5px}
  	\includegraphics[width=1.0\linewidth]{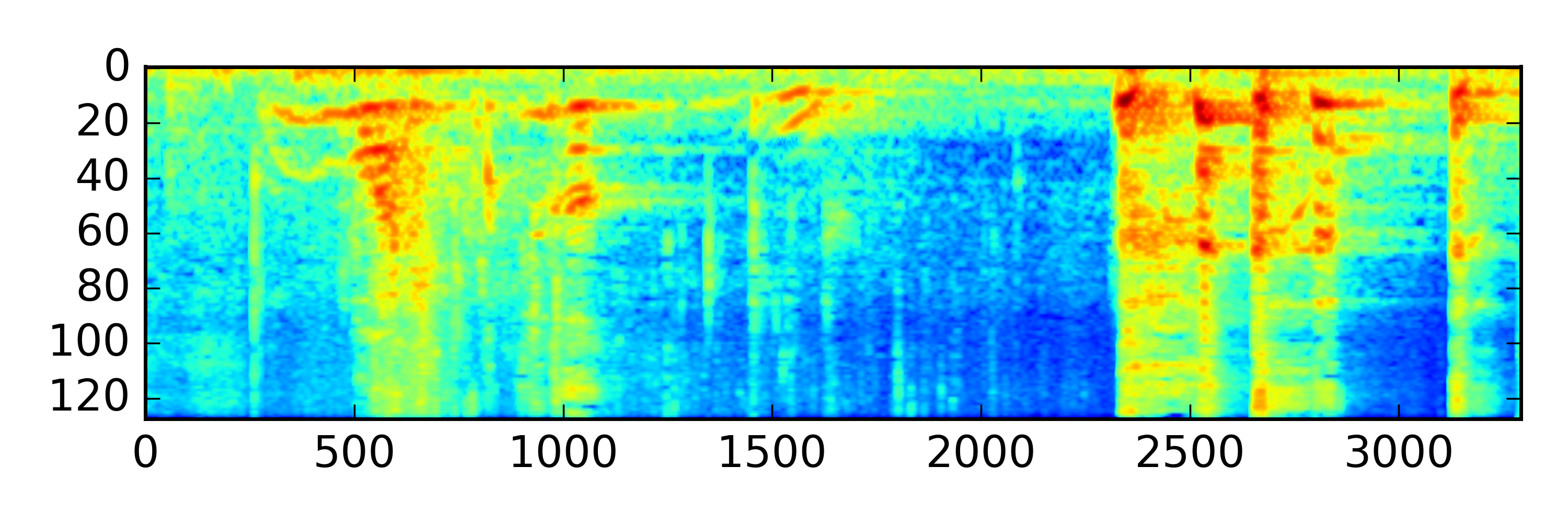}
  \end{minipage}
  &
  \begin{minipage}{.47\textwidth}
  	\vspace{5px} 
  	\includegraphics[width=1.0\linewidth]{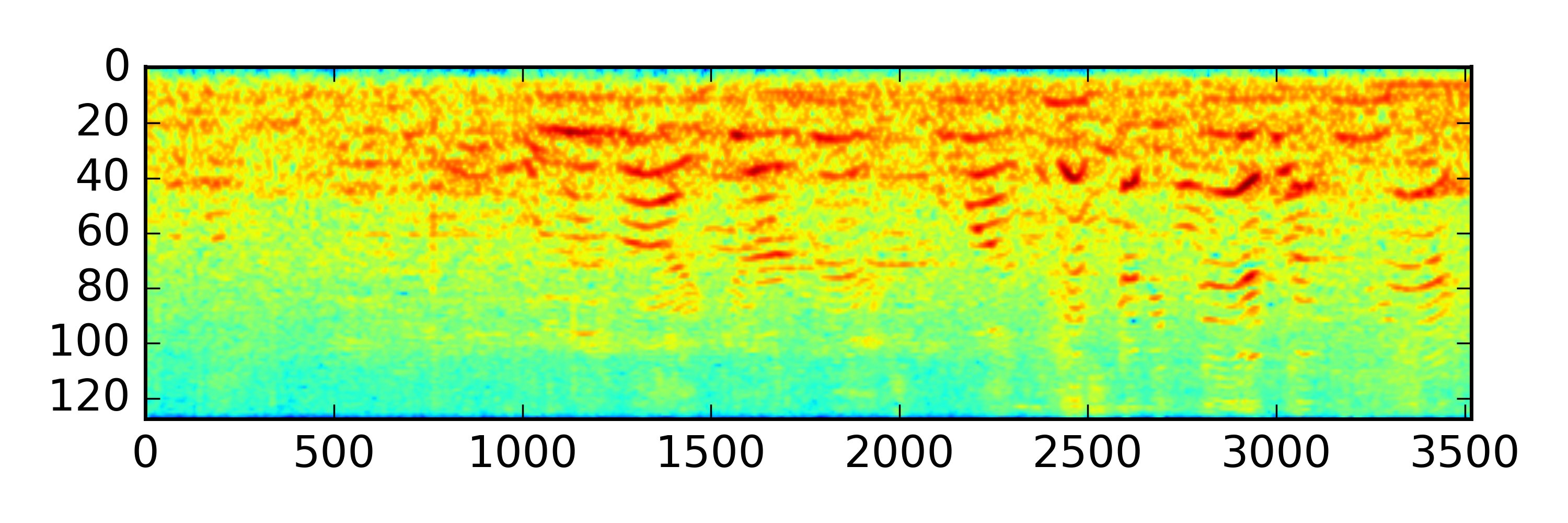}
  \end{minipage}\\
  
  \hline
  
  \begin{minipage}{.47\textwidth}
  	\vspace{5px}
  	\includegraphics[width=1.0\linewidth]{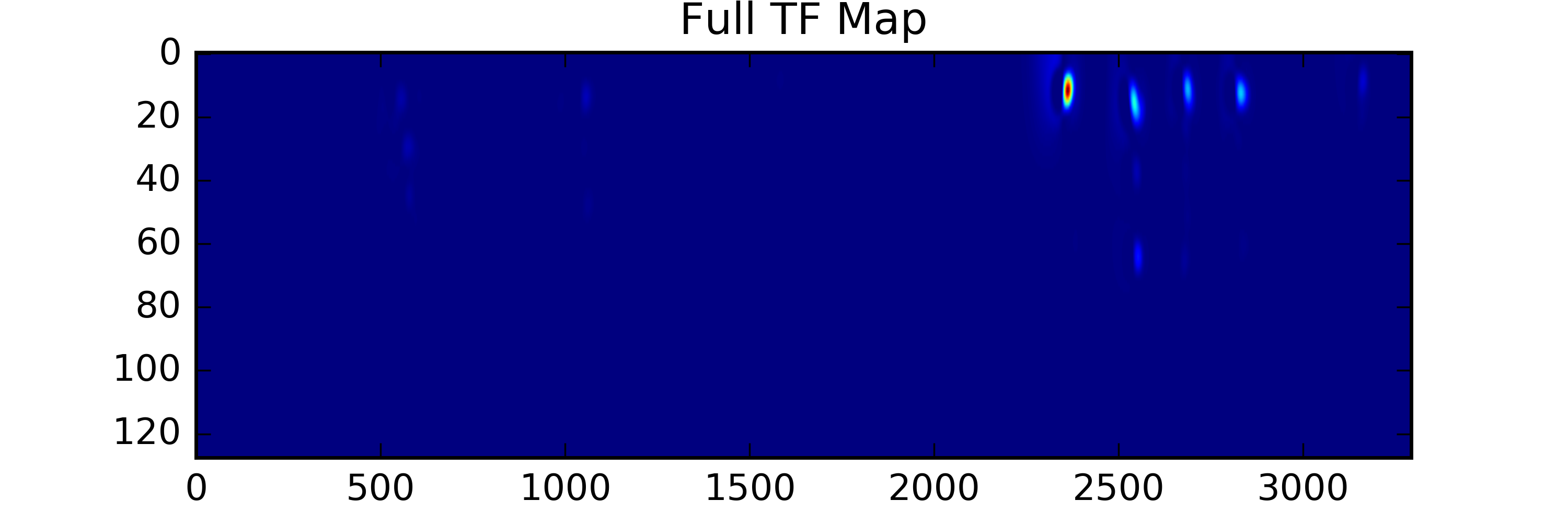}
  \end{minipage}
  &
  \begin{minipage}{.47\textwidth}
  	\vspace{5px}
  	\includegraphics[width=1.0\linewidth]{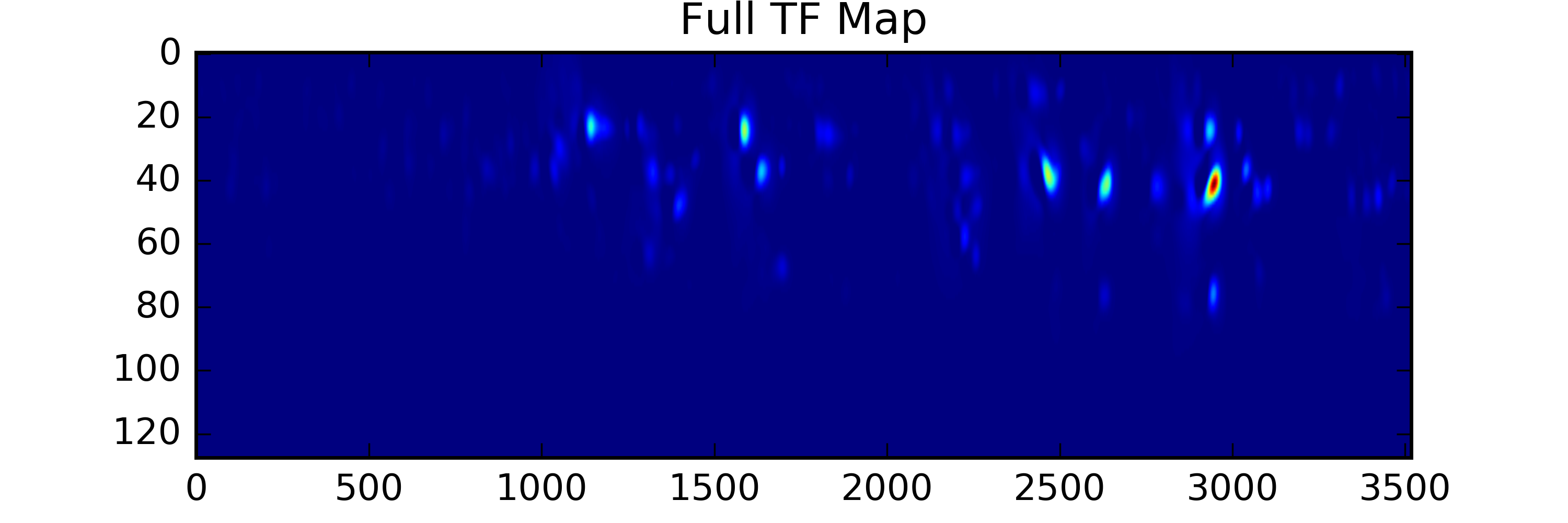}
  \end{minipage}\\
  
  \hline
  
  \begin{minipage}{.47\textwidth}
  	\vspace{5px}
  	\includegraphics[width=1.0\linewidth]{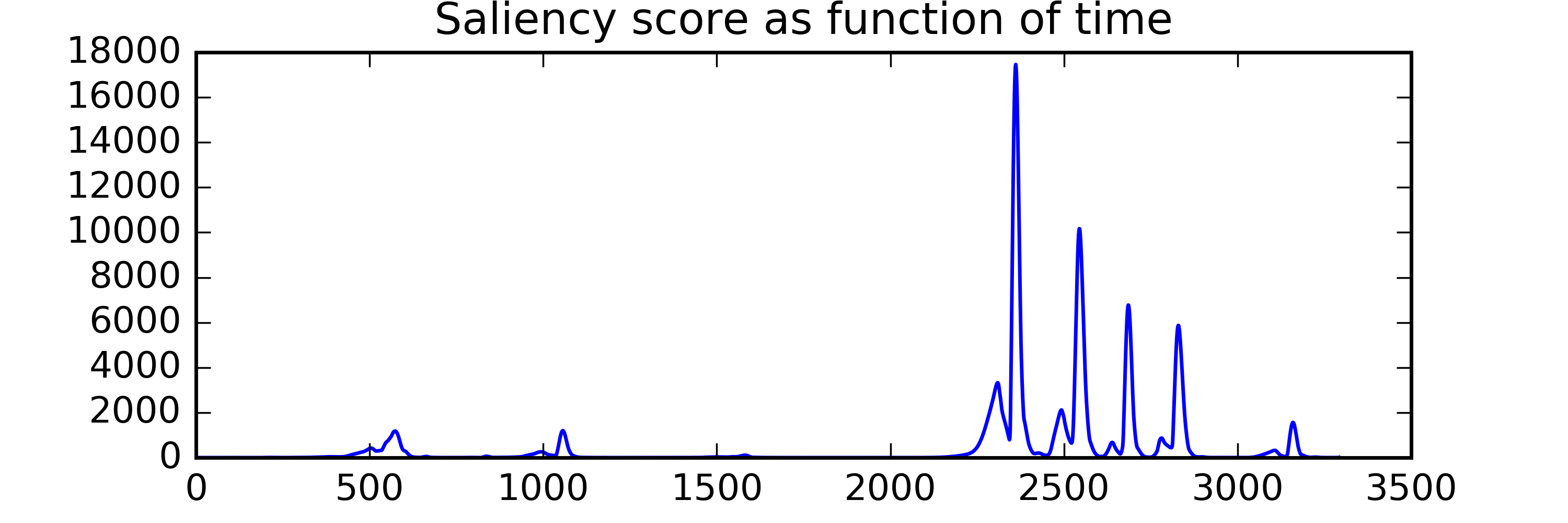}
  \end{minipage}
  &
  \begin{minipage}{.47\textwidth}
  	\vspace{5px}
  	\includegraphics[width=1.0\linewidth]{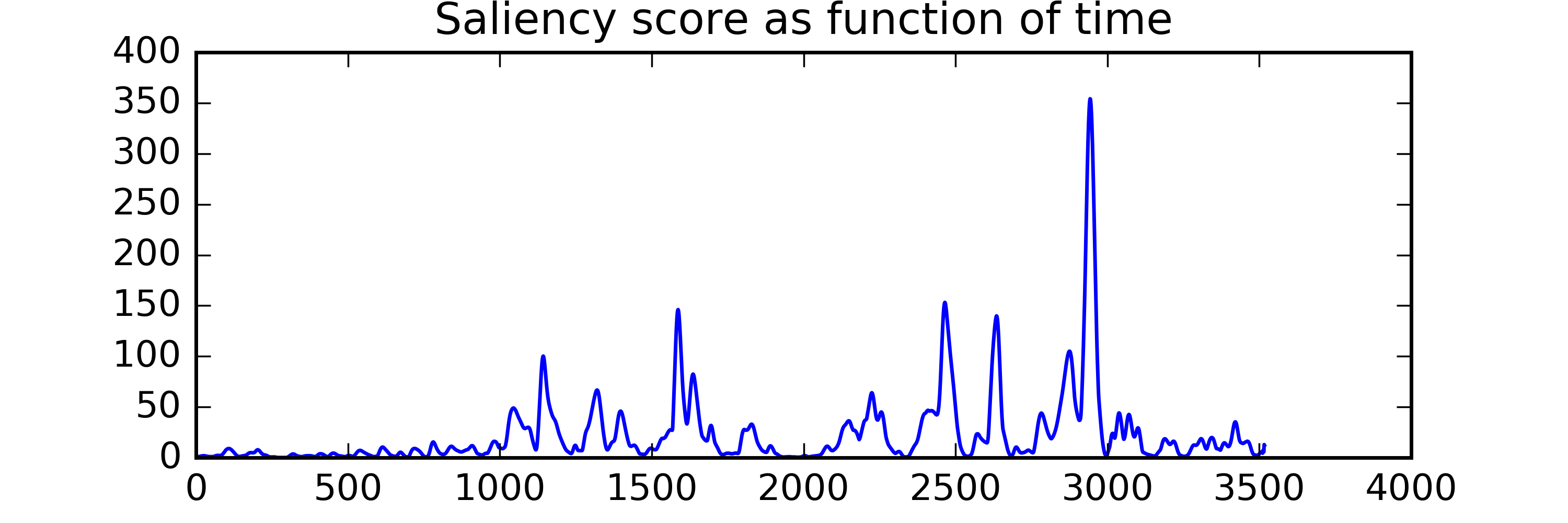}
  \end{minipage}\\
  
  \hline
  
  \begin{minipage}{.47\textwidth}
  	\vspace{5px}
  	\includegraphics[width=1.0\linewidth]{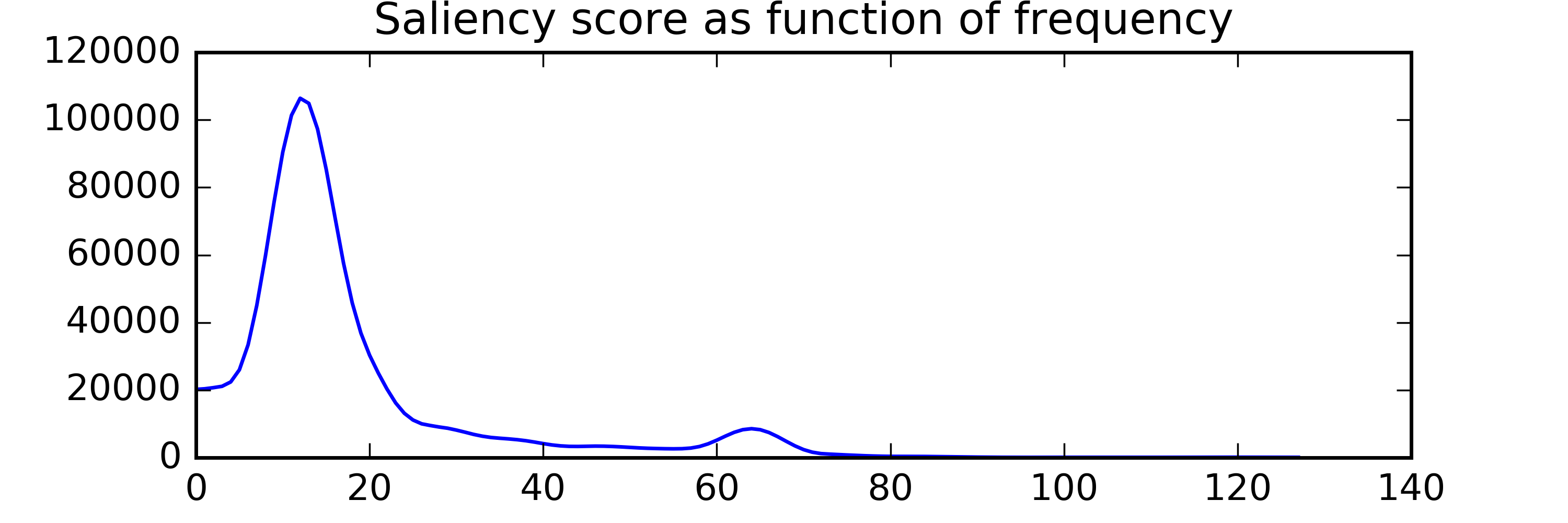}
  \end{minipage}
  &
  \begin{minipage}{.47\textwidth}
  	\vspace{5px}
  	\includegraphics[width=1.0\linewidth]{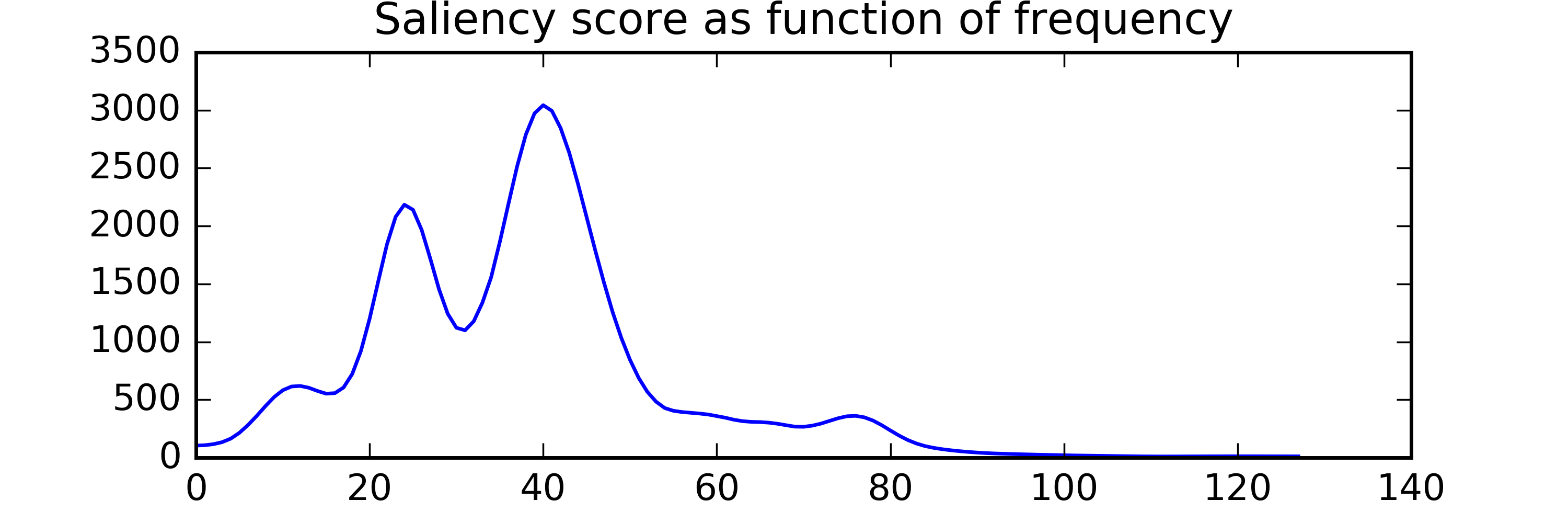}
  \end{minipage}\\
  
  \hline
  
\end{tabular}
}
\caption{A table demonstrating the auditory salience model based on \cite{kayser2005mechanisms} applied to two contrasting audio samples in the HCU400 dataset. The resulting salience scores (bottom) are summarized and used as features in predicting memorability.}
\label{salience}
\end{figure*}

\section{Samples and Feature Generation}

Audio samples for this test were taken from the HCU400 dataset \cite{OURSELVES}.  Standard low-level acoustic features were extracted from each sample based on prior precedent \cite{richard2013overview}.  We used default configurations from three audio analysis tools: Librosa \cite{librosa}, pyAudioAnalysis \cite{pyaudioanalysis}, and Audio Commons \cite{audiocommons}, which include basic features (i.e. spectral spread) as well as more advanced timbral modeling.  We supplement these features with additional summary statistics like high/mid/bass energy ratios, percussive vs. harmonic energy, and pitch contour diversity.

Over the last decade there have been advances in cognitive models that can determine the acoustic salience of sound, inspired by the neuroscience of perception \cite{delmotte2012computational,kalinli2009saliency}.  Here we follow the procedure proposed by \cite{kayser2005mechanisms}, applying separate temporal, frequency, and intensity kernels to an input magnitude spectrogram to produce three time-frequency salience maps.  Figure \ref{salience} shows a comparison of temporal salience between two sound samples in the HCU400 dataset with highly contrasting auditory properties.  From these maps, we compute a series of summary statistics to be used as features.

High-level, top-down features were taken from our previous work in \cite{OURSELVES} and include causal uncertainty ($H_{cu}$), the cluster diameter of embedding vectors generated from user-provided labels (quantifying source agreement or source location), familiarity, imageability, valence, and arousal.  

\section{Measuring Memorability}

\begin{figure}[h!]
\includegraphics[width=8cm]{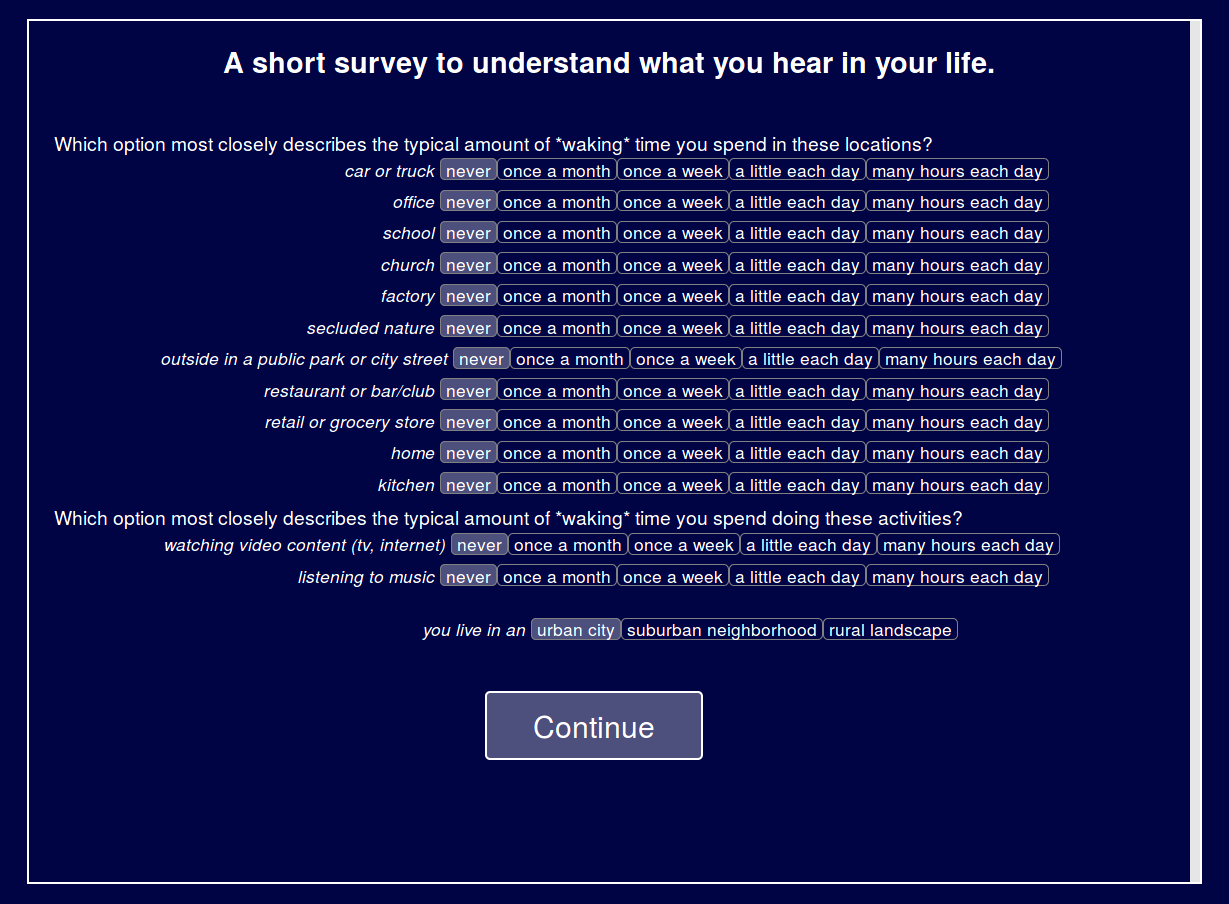}
\vspace{2px}
\includegraphics[width=8cm]{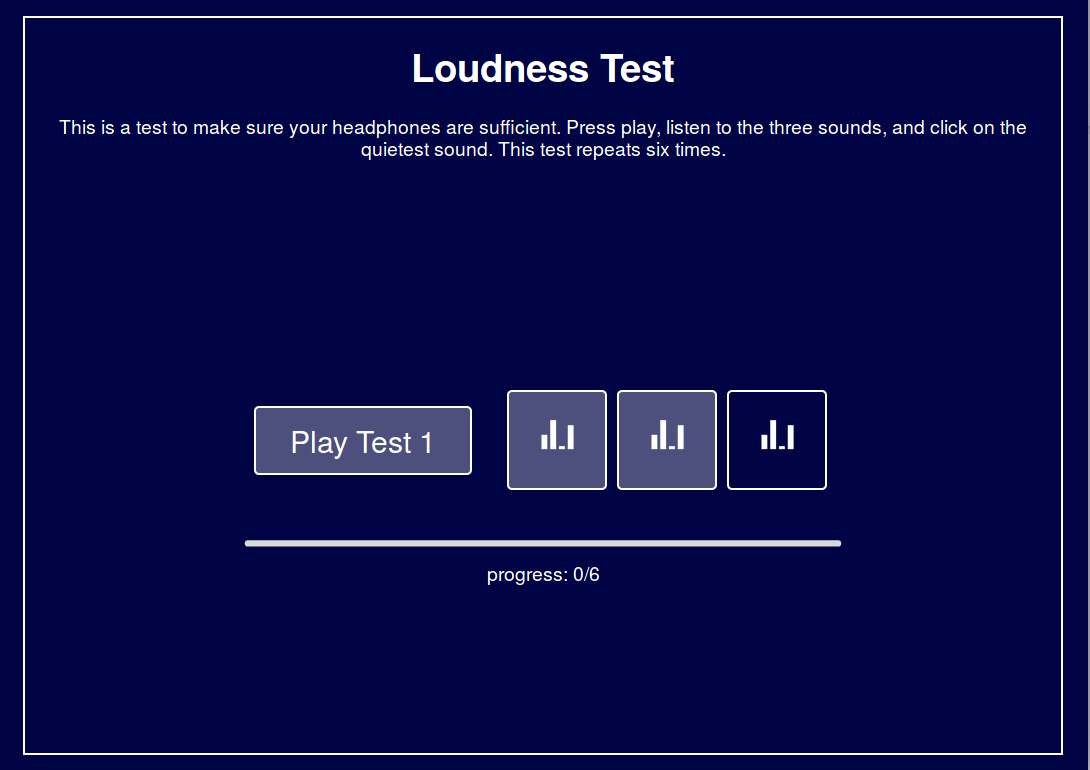}
\vspace{2px}
\includegraphics[width=8cm]{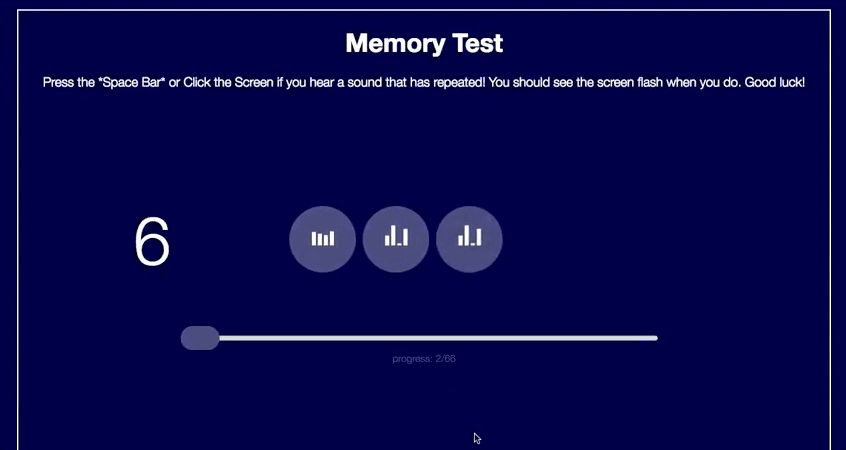}
\caption{Screenshots of the auditory memory game interface presented to participants as a part of our study.  The game can be found at \url{http://keyword.media.mit.edu/memory.}}
\label{memtest}
\end{figure}

In order to quantify memorability, we drew inspiration from work in \cite{faces}, which used an online memory game to determine the features that make images memorable.  We designed an analogous interface for the audio samples in the HCU400 dataset; this interface can be found at \url{http://keyword.media.mit.edu/memory}. The game opens with a short auditory phase alignment-based assessment \cite{headphones} to ensure that participants are wearing headphones, followed by a survey that captures data about where they spend their time (urban vs. rural areas, the workplace vs home, etc).  Participants are then presented with a series of 5 second sound clips from the HCU400 dataset, and are asked to click when they encounter a sound that they've heard previously in the task.  At the end of each round consisting of roughly 70 sound clips, the participant is provided with a score. Screenshots of the interface at each stage are shown in Figure \ref{memtest}. 

By design, each round of the game consisted of 1-2 pairs of \textit{target sounds} and 20 pairs of \textit{vigilance sounds}.  \textit{Target sounds} were defined as samples from the dataset that were separated by exactly 60 samples-- the sounds for which memorability was being assessed in a given round.  The \textit{vigilance sounds}, pairs of sounds that were separated in the stream by 2 to 3 others, were used to ensure reliable engagement throughout the task following the method in \cite{faces}.  Roughly 20,000 samples were crowd-sourced on Amazon Mechanical Turk such that a single task consisted of a single round in the game.  Individual workers were limited to no more than 8 rounds to ensure that target samples were not repeated.  Rounds that failed to meet a minimum vigilance score (>60\%) or exceeded a maximum false positive rate (>40\%) were discarded.

\section{Summary of Participant Data}
We recruited 4488 participants, consisting of a small ($<$50) number of volunteers from the university community and the rest from Amazon Mechanical Turk.  Our survey data shows that our participants report a 51/37/12\% split between urban, suburban, and rural communities.  We see weak trends in the average time per location reported for each community type-- urbanites self-report spending less time at home, in the kitchen, in cars, and watching media on average.  Rural participants report spending more time in churches and in nature.  Using KNN clustering and silhouette analysis, we find four latent clusters -- students (590 users), office workers (1250 users), home-makers (1640 users), and none of these (1010 users).  Split-rank comparisons between groups did not reveal meaningful differences in results across user groups; we speculate any differences due to ecological exposure of sounds between environments is not consistent or influential enough at this group level to alter performance. 

\begin{figure}[h!]
\centering
\includegraphics[width=6cm]{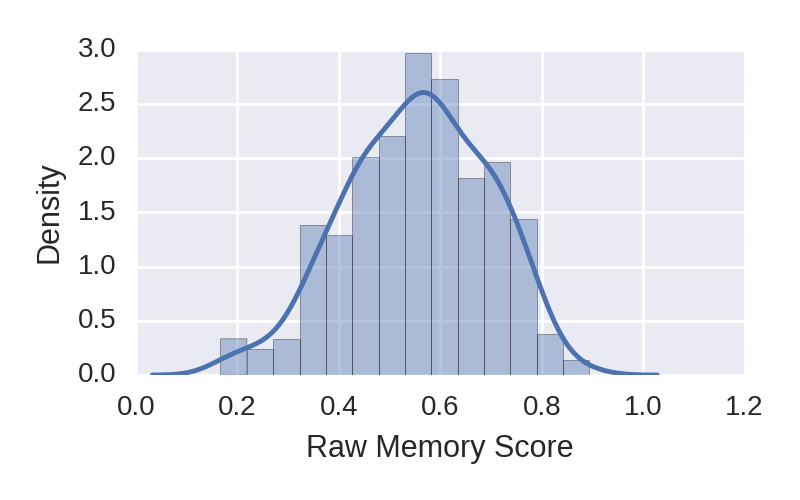}
\includegraphics[width=6cm]{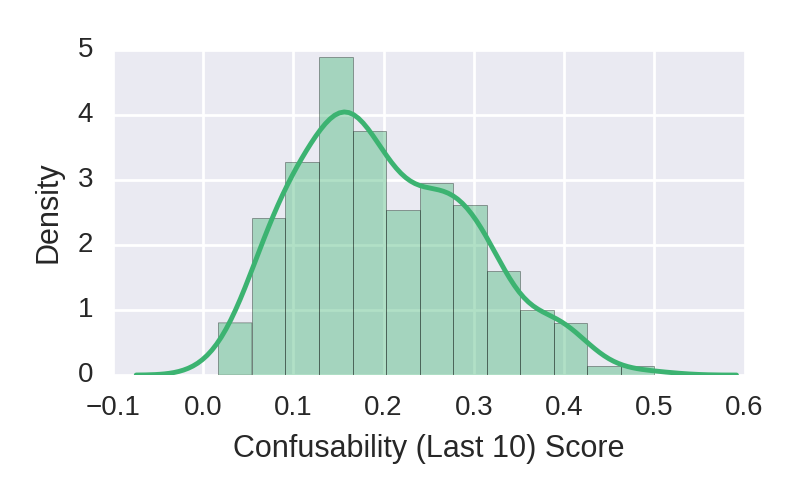}
\caption{Top: A histogram of the raw scores for each sound -- they were successfully remembered and identified about $55\%$ of the time on average, with a large standard deviation; Bottom: A histogram of "confusability" scores for each sound, with an average score of about $25\%$.}
\label{hist_results}
\end{figure}

\section{Summary of Memory Data} 
The raw memorability score $M$ for each sound is simply computed as the number of times it was correctly identified as the target divided by the number of its appearances.  However, this does not account for the likelihood that the sound will be falsely remembered (i.e. clicked on without a prior presentation). We additionally compute a "confusability" score $C_{10}$ for each sound sample, defined as the false positive rate for sounds when they fall close to the second target presentation (i.e. in the last ten positions of the game).  We can thus derive a "normalized memory score" represented by $M - C_{10}$.  In attempting to understand auditory memory, we consider both what makes a sound memorable \textit{and} what makes a sound easily mis-attributed to other sounds, whether those sounds are encountered in our game or represent the broader set of sounds that one encounters on a habitual basis.  We therefore model both normalized memorability and confusability in this work.

\begin{figure}[h!]
    \centering
	\includegraphics[width=6cm]{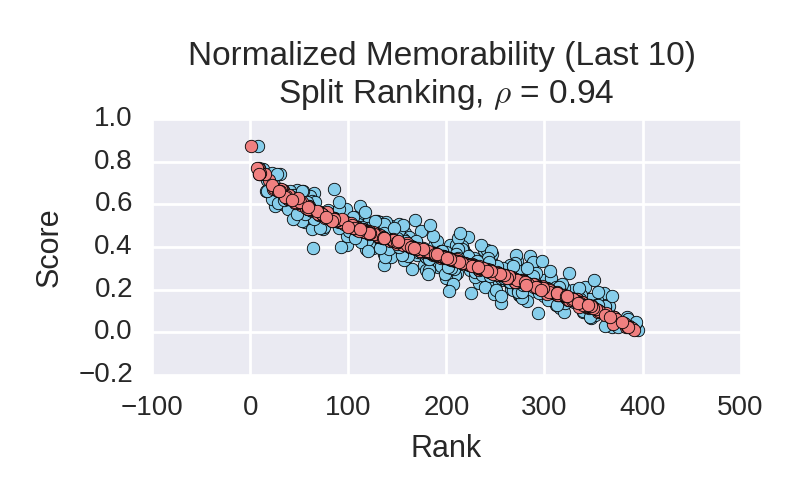}
    \includegraphics[width=6cm]{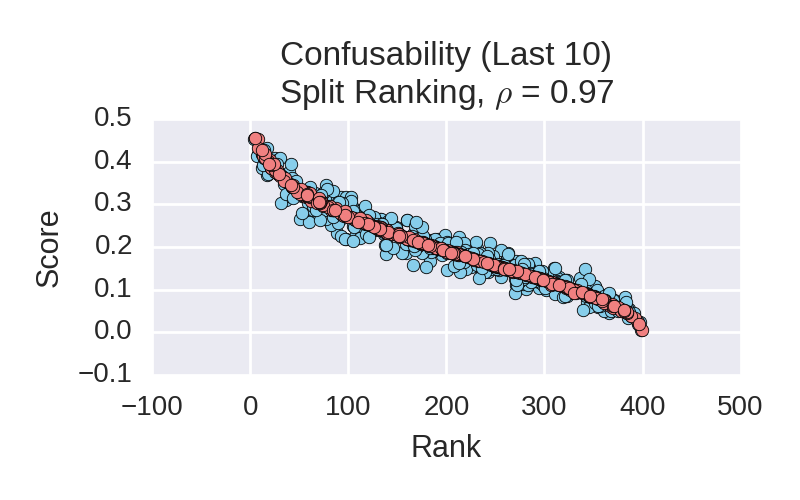}
    \caption{The results of the split-ranking analysis for the normalized memorability score and confusability score, using 5 splits; The Spearman coefficient correlations demonstrate the reliability of these scores across study participants, enabling us to model both metrics in the later parts of the work.}
    \label{mem reliability}
\end{figure}

We confirm the reliability of both the normalized memory scores and the confusability scores across participants by performing a split ranking analysis similar to \cite{faces} with 5 splits, shown in Figure \ref{mem reliability} with their respective Spearman correlation coefficients.  This confirms that memorability and confusability are consistent, user-independent properties.

In Table \ref{memorable sounds}, we show a short list of the most and least memorable and confusable sounds in our dataset as a function of the normalized memorability score and confusability score.

\begin{table}[h!]
\resizebox{\columnwidth}{!}{
\begin{tabular}{|l|l|}
\hline
\textbf{Most Memorable} & \textbf{Least Memorable}\\
\hline
\textit{man\_screaming.wav} & \textit{morphed\_firecracker\_fx.wav}\\
\textit{woman\_screaming.wav} & \textit{truck\_(idling).wav}\\
\textit{flute.wav }& \textit{morphed\_turkey2\_fx.wav}\\
\textit{woman\_crying.wav} & \textit{morphed\_airplane\_fx.wav}\\
\textit{opera.wav} & \textit{morphed\_metal\_gate\_fx.wav}\\
\textit{yawn.wav} & \textit{morphed\_shovel\_fx.wav}\\
\hline
\textbf{Most Confusable} & \textbf{Least Confusable}\\
\hline
\textit{garage\_opener.wav} & \textit{clock.wav}\\
\textit{lawn\_mower.wav} & \textit{morphed\_335538\_fx.wav}\\
\textit{washing\_machine.wav }& \textit{phone\_ring.wav}\\
\textit{rain.wav} & \textit{woman\_crying.wav}\\
\textit{morphed\_tank\_fx.wav} & \textit{woman\_screaming.wav}\\
\textit{morphed\_printing\_press\_fx.wav} & \textit{vomit.wav}\\
\hline
\end{tabular}
}
\caption{A list of the most and least memorable and confusable sounds from the HCU400 dataset.}
\label{memorable sounds}
\end{table}

\section{Feature Trends in Memorability and Confusability}
We consider two objectives -- (1) to determine the relationship between individual features and our measured memorability and confusability scores, and (2) to determine the relative importance of these features in predicting memorability and confusability.  To address the former, we provide the resulting $R^2$ value after applying a transform learned using support vector regression (SVR) for each individual feature.  For the latter, we use a sampled Shapely value regression technique in the context of SVR-- that is, we first take $N$ random features ($N$ between 1 and 10), perform an SVR to predict memorability or confusability scores for our 402 sounds and the calculated $R^2$ of the fit.  We then measure the change in $R^2$ as we append every remaining feature to the model, each individually.  The largest average changes over 10k models are reported in table \ref{Shapely}.  This technique is robust to complex underlying nonlinear relationships from feature space to predicted metric as well as feature collinearity. 
We find that the strongest predictors of both memorability and confusability are the measures of imageability (how easy the sound is to visualize) and its causal uncertainty.  Memorability is dominated by high level, gestalt features, with only one lower level feature (`pitch diversity') in the ten most important features.  Low level features, including those derived from the auditory salience models, play a more significant role in determining confusability.  

\begin{figure*}[h!]
\centering
\includegraphics[width=15cm]{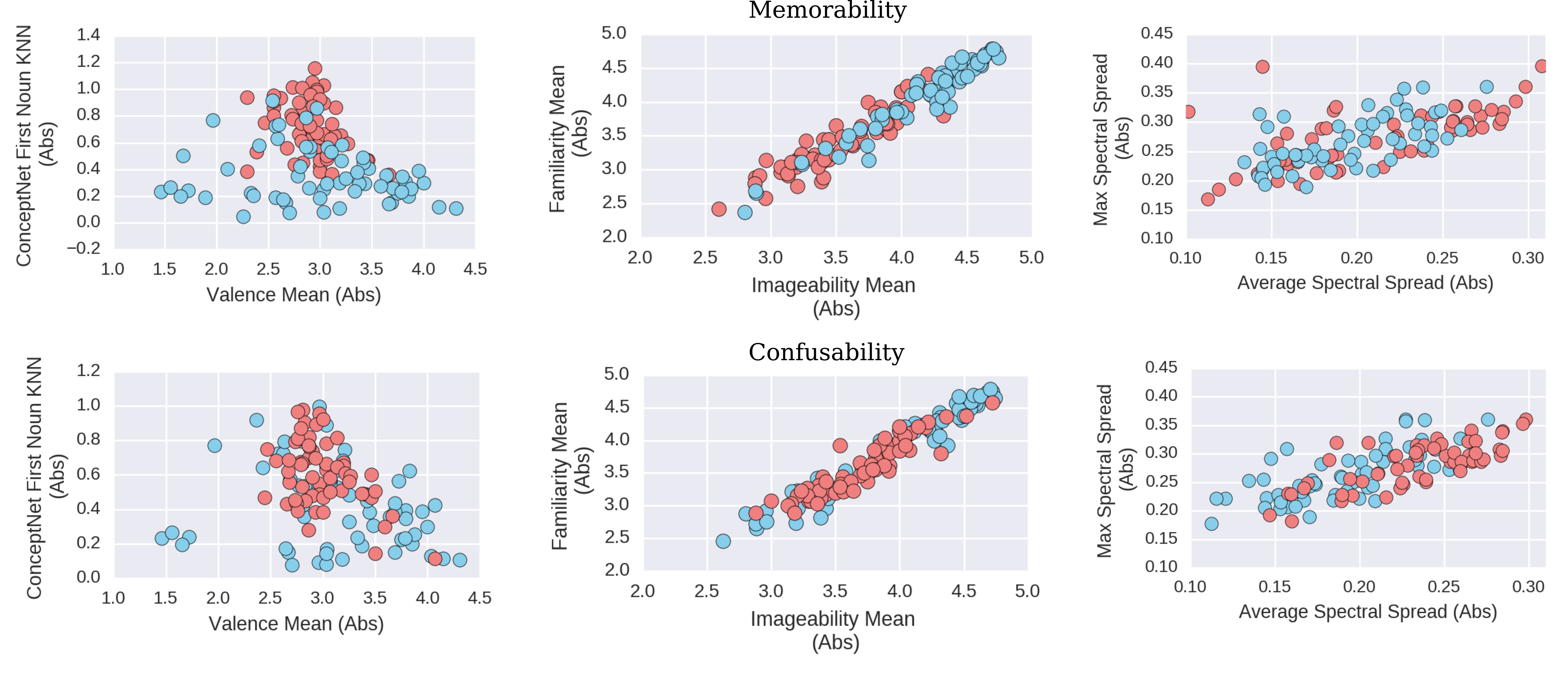}
\caption{Scatter plots showing the changes in distribution of select features based on extremes in  memorability (top row) and confusability (bottom row); blue indicates sounds that are most (85th percentile) memorable or least (15th percentile) confusable; red indicates sounds that are least memorable or most confusable.}
\label{scatter}
\end{figure*}

The absolute $R^2$ values indicate that no individual feature is a significant predictor of memorability by itself. This implies a complex causal interplay in feature space, which we explore further in the set of plots presented by Figure \ref{scatter}.  In each plot, we show a distribution of feature values for the 15\% of sounds that are most memorable or least confusable (blue) contrasted against the least memorable or most confusable sounds (red).  We first consider the effect of $Hcu$ and valence on memory-- low memorability and high confusability sounds exhibit a similar trend of high causal uncertainty and neutral valence (Column 1). \cmmnt{; however, while a fairly broad distribution of valence drives high memorability, positive valence (eliciting excitement or anxiety) tends to be a much stronger driver of low confusability.} In Column 2, we consider imageability and familiarity ratings, shown to be strongly collinear in \cite{OURSELVES}.  Here, their relationship to memorability and confusability diverge; while both are positively correlated with memorability, \textit{neutral} ratings are the stronger predictor of confusability.  This suggests that we are most likely to confuse sounds if they are loosely familiar but neither strictly novel nor immediately recognizable.  Finally, Column 3 reveals a discernible decision boundary in low-level feature space for confusability which doesn't exist in its memorability counterpart.  The relative importance of low-level salience features, here represented by spectral spread, aligns with intuition-- we hypothesize that, in the absence of strong causal uncertainty or affect feature values, our perception of sounds is driven by their spectral properties.  

\begin{table*}[h!]
\centering
\resizebox{.8\textwidth}{!}{
\begin{tabular}{|l|r|r|l|r|r|}
\hline
					 \multicolumn{6}{|c|}{Top Predictors for Memorability and Confusability} \\
\hline
\multicolumn{3}{|c|}{Memorability} & \multicolumn{3}{|c|}{Confusability} \\
\hline
Feature & $R^2$ & Shapely $\Delta$ $R^2$ & Feature & $R^2$ & Shapely $\Delta$ $R^2$ \\
\hline

\textbf{Imageability} & 0.201 & 0.126 & \textbf{Imageability} & 0.065 & 0.078\\
\textbf{Hcu} & 0.224 & 0.125 & \textbf{Hcu} & 0.073 & 0.078\\
\textbf{Familiarity} & 0.176 & 0.123 & Avg Spectral Spread & 0.087 & 0.078\\ 
\textbf{Valence} & 0.178 & 0.120 & Peak Spectral Spread & 0.037 & 0.076\\
\textbf{Location Embedding Density} & 0.147 & 0.117 & Peak Energy, Frequency Salience Map & 0.059 & 0.076\\
\textbf{Familiarity std} & 0.103 & 0.117 & \textbf{Location Embedding Density} & 0.100 & 0.076\\ 
Pitch Diversity & 0.084 & 0.113 & Frequency Skew, Frequency Salience Map & 0.059 & 0.076\\ 
\textbf{Imageability std} & 0.086 & 0.113 & \textbf{Arousal} & 0.039 & 0.076\\ 
\textbf{Arousal} & 0.072 & 0.112 & Peak Energy, Intensity Salience Map & 0.044 & 0.075\\
\textbf{Arousal std} & 0.056 & 0.111 & \textbf{Familiarity} & 0.045 & 0.075\\

\rule{0pt}{2.5ex}\textit{Avg Spectral Spread} & \textit{0.099} & \textit{0.107} & \textit{\textbf{Valence}} & \textit{0.100} & \textit{0.075}\\
\textit{Timbral Sharpness} & \textit{0.094} & \textit{0.091} & \textit{Timbral Roughness} & \textit{0.094} & \textit{0.047}\\
\textit{Max Energy} & \textit{0.091} & \textit{0.100} & \textit{Avg Flux, Sub-band 1} & \textit{0.092} & \textit{0.064}\\
\textit{Treble Energy Ratio} & \textit{0.090} & \textit{0.020} & \textit{Flux Entropy, Sub-band 1} & \textit{0.091} & \textit{0.061}\\ 


\hline
\end{tabular}
}
\caption{The top performing features from the Shapely regression analysis for both memorability and confusability (gestalt features are bolded); shown are the features ordered by their respective contributions to the $R^{2}$ value, with additional features with top performing individual $R^2$ values appended in italics.  The first column indicates the individual predictive power of each feature; the second indicates its relative importance in the context of the full feature set.}
\label{Shapely}
\end{table*}



\section{Per-game Modeling of Memorability}

The aural context in which a sound is presented, which includes ecological exposure as well as the immediate preceding sounds in our audition task, may influence the memory formation process.  The literature supports the notion that, given a context, unexpected sounds are more likely to grab our attention and engage memory \cite{schirmer2011perceptual}.  To understand this effect in our test, we ran two studies based on a 5 sound context (approximating the limits of semantic working memory) and a 1 sound context (approximating the limits of echoic memory).  

Table \ref{logistic} shows the results of a model trained to predict whether the target in each game will be successfully recalled.  This model was trained with the most memorable and least memorable sounds only (15th/85th percentiles) with a 5-fold cross-validation process, and results are reported on a 15\% hold-out test set.

To begin, a baseline model is trained using the absolute, immutable features of the target sound.  Because there are a limited number of sounds in our dataset relative to the number of games, the feature space is redundant and sparse, and we expect the accuracy of this model to converge to the average expected value over our set of sounds.  We then introduce contextual features-- the \textit{relative difference} (z-score) of target sound features with those of the varying sounds that precede its first presentation in each game-- to see if our model improves.  Evidence of over-fitting on the train and validation set when all of the contextual features are included (decreasing test set accuracy) motivates a second test, limited to a smaller set of the 50 most meaningful features (from our SVR analysis; 25 high-level and 25 low-level).  In both cases, however, model performance does not improve as we would expect if the context contained additional useful information.

We also run a classifier that \textit{only} uses contextual features, to ensure informative context has not been obscured, as useful information in the context could be subsumed by the absolute features in our first test.  We run a noise baseline in which contextual features are calculated using a random context, which are still informative as the z-score depends largely on the absolute features of the target sound.  We then run a model with the proper context to measure the difference in performance.  There is no improvement when the proper context is re-introduced.     

This leads us to a meaningful insight, contrary to our hypothesis -- context does \textit{not} exert a measurable influence on our results.  While context likely does matter in real-world settings, we suspect that our memory game framework indirectly primes participants to expect otherwise surprising sounds.  This confirms that our data is the consequence of truly intrinsic properties of the sounds themselves, independent of immediate context \textit{and} participant ecological exposure (as was demonstrated in the split-rank analysis).

\begin{table}[h!]
\resizebox{\columnwidth}{!}{
\begin{tabular}{|l|r|}
\hline
					 \multicolumn{2}{|c|}{Memorability Per-Game Models} \\
\hline
\textbf{Features} & \textbf{Accuracy (\%)}\\
\hline
Absolute + All 5-Sound Context Feats (working semantic) & 68.0 \\
Absolute + Top 50 5-Sound Context Feats & 69.1 \\
\textit{Absolute Feature Only Baseline (\textasciitilde expected value)} & \textit{70.3} \\

\rule{0pt}{2.75ex}Contextual Only, 5-Sound Context (working semantic) & 62.5 \\
\textit{5 Sound Context, Noise Baseline} & \textit{64.1} \\
\hline
\rule{0pt}{2.75ex}Absolute + All 1-Sound Context Feats (echoic) & 68.0 \\
Absolute + Top 50 1-Sound Context Feats &  69.5 \\
\textit{Absolute Feature Only Baseline (\textasciitilde expected value)} & \textit{70.3} \\

\rule{0pt}{2.75ex}Contextual Only, 1-Sound Context (echoic) & 60.0 \\
\textit{1 Sound Context, Noise Baseline} & \textit{61.3} \\

\hline
\end{tabular}
}
\caption{The influence of contextual sounds before the first presentation of the target on our ability to predict recall across games.}
\label{logistic}
\end{table}



\section{Implications and Conclusion}

In this work, we quantify the inherent likelihood that a sound will be remembered or incorrectly confused and confirm that is consistent across user groups. In line with our hypotheses, we show that the most important features that contribute to a sound being remembered are gestalt-- namely those sounds with clear sound sources (high $H_{cu}$), that are easy to visualize, familiar, and emotional. We also show that low $H_{cu}$ sounds that are not familiar or easy to visualize are most likely to be mis-attributed, and low level features play a more important role in predicting this behavior.  These relationships are not influenced by context, and are intrinsic properties of the sounds themselves.

To our knowledge, this is the first body of work that combines top-down theories from psychology and cognition with bottom-up auditory salience frameworks to model the memorability of everyday sounds. We posit that the demonstration of memorability as an intrinsic, user and context-independent property of sounds, along with the insights mentioned above, have significant implications for audio technology research -- for example, knowing that gestalt features are the primary drivers of memorability might allow us to selectively choose audio samples in a stream or sound environment to be recorded and stored, as a way of mimicking human memory to perform compression at a level of abstraction higher than the sample level.  An understanding of the most significant predictors of memorability and confusability might also allow us to artificially manipulate our sonic environments to make certain streams of audio more or less memorable, perhaps as a memory aid or a mechanism to eliminate distractions vying for our attention.  Looking ahead, we aim to enable many of these applications by translating the principles from this work to an online, real-time model.

\bibliography{aes2e.bib}
\bibliographystyle{jaes.bst}

\end{document}